# Ultralong trapping of light using double spin coherence gratings

B. S. Ham[*] and J. Hahn

Center for Photon Information Processing, and the Graduate School of Information and Telecommunications,
Inha University, 253 Yonghyun-dong, Nam-gu, Incheon 402-751, S. Korea
[*]*Corresponding author: bham@inha.ac.kr*

**Abstract:** Ultralong trapping of light has been observed in an optically dense three-level solid medium interacting with a pair of counterpropagating coupling fields. Unlike the light trapping based on standing-wave gratings excited by the same frequency pair of counterpropagating light fields (M. Bajcsy *et al.*, *Nature* **426**, 638 (2003)), the present method uses resonant Raman optical field-excited spin coherence gratings. The observed light trapping time is two orders of magnitude longer than the expected value of the spin dephasing time, where the extended storage time has potential for quantum information processing based on nonlinear optics.

In this Letter, we experimentally demonstrate an ultralong trapping of light (stationary light) in a double electromagnetically-induced-transparency (EIT) scheme of a three-level solid medium, whose spin transition is inhomogeneously broadened. Recently light-matter interactions have drawn much attention to quantum optics since the observations of ultraslow light in a Bose Einstein condensate [1]. The physics of group velocity control of a traveling light lies in the Kramers-Kronig relation, where the absorption spectrum correlates with a dispersion line shape [2]. This absorption-dispersion relation leads to a refractive index control of an optical medium. Ultraslow light has been observed in atomic media [1,3], solid states [4,5], semiconductors [6], photonic crystals [7], and optical fibers [8]. Even ultraslow-light-based stationary light [9] has been demonstrated by using standing-wave gratings based on EIT [10] in atomic vapors. On the other hand, quantum mapping processes between photons and atomic spins have been demonstrated for a light storage process in quantum memory applications [4,11-13]. Unlike the quantum mapping process, both ultraslow light and light stopping processes contribute to nonlinear quantum optics in the context of lengthening interaction time with an optical medium to induce a giant phase shift even with ultralow power light [9,14-16]. In conventional nonlinear optics such as cross phase modulation, use of intense light poses a critical limit. Thus, the giant phase shift of weak signal light has potential applications not only in nonlinear optics such as electro-optic switches [17], but also in quantum information processing such as entanglement generation [18-20] and quantum nondemolition measurement [21].

In the quantum mapping process (zero group velocity), however, the light component must be converted into spin coherence, and thus the storage time should be limited by the spin dephasing time of the medium [4,9,11-13]. Although spin coherence is robust and more stable compared to an optical counterpart, spin inhomogeneity makes the spin dephasing time much shorter in most solid media. The drawback of shorter spin dephasing time, however, can be overcome by using a reversible echo-type coherence process [4,22,23], where the stopping time becomes as long as the spin coherence decay time in a homogeneous system [11-13]. The quantum mapping process, however, prohibits nonlinear interactions because the signal light must be absorbed into the medium during the stopping period. The detailed work of quantum optical data storage using the spin rephrasing process in a spin inhomogeneous system has been investigated recently [4,22,23].

The observed light trapping in this Letter is fundamentally different from the previous light stopping phenomenon based on the standing-wave grating [9] or quantum mapping process [4,11-13]. The physics of the present light trapping mechanism is in the mutual interactions of oppositely moving spin coherence gratings [24]. Very recently we have discussed that EIT-based optical storage or simply the quantum mapping process is equal to the process of coherence conversion from photons to atomic spins and vice versa [25], where spin coherence acts as moving gratings which result from resonant Raman field excitations [22].

Figure 1(a) shows a partial energy-level diagram of a rare-earth $Pr^{3+}$ (0.05 at. %) doped $Y_2SiO_5$ (Pr:YSO) [26]. Pr:YSO is a persistent spectral hole burning medium, where the spin population decay time $T^S_1$ among the ground states (|1>, |2>, and |3>) is much longer ($T^S_1 \sim$ minute) than the optical population decay time $T^O_1$ between the ground and excited states ($T^O_1 \sim 110$ μs) [26]. Thus, the light R whose pulse duration is 10 ms acts as a repump beam functioning atom transfer from state |0> to state |1> for the purpose of (i) increasing the atom density in state |1>, and (ii) maintaining the same initial condition for the repeated probe light. So does C if it precedes P. The resonant probe pulse P can experience an absorption cancellation under the coupling light C owing to EIT. The field A, whose propagation direction is opposite to that of the coupling C, is used to generate backward spin coherence moving gratings along with phase conjugate, PC, whose direction is opposite to that of P according to the phase matching condition: $\mathbf{k}_{PC} = \mathbf{k}_C − \mathbf{k}_P + \mathbf{k}_A$. The light Y is used to prove the spin coherence conversion process, so-called photon switching [27]. The existence of the diffracted signal D is a direct proof of the photon switching (discussed in Fig. 2(c)).

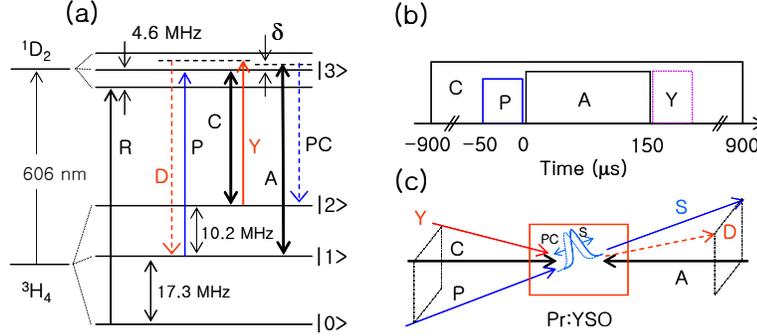

**Fig. 1.** (Color online) (a) A schematic of partial energy levels of $Pr^{3+}$ doped $Y_2SiO_5$, (b) pulse sequence, and (c) light propagation scheme. The light D satisfies phase matching condition with C, P, and Y. Y is applied for only Fig. 2(c), which is 1 MHz detuned from C. $\delta$ = 20 kHz.

Figures 1(b) and 1(c) show schematics of pulse sequence and propagation directions, respectively. The repump beam R precedes the coupling light C (not shown). The pulse train of Fig. 1(b) is repeated at 20 Hz. The angle between the light beams P, C, R, and Y is ~25 mrad, where the beams remain spatially overlapped overall at 80% inside the sample of Pr:YSO. All light beams are focused by a lens. The beam spot diameter (exp(-1) in intensity)) of the P at a focal point is ~200 μm; others are ~600 μm. The sample of Pr:YSO is in a helium cryostat keeping temperature at 5 K. Because light-matter interaction time increases in the ultraslow light regime, the spin coherence excitation must be enhanced. Thus, the nondgenerate four-wave mixing process must be efficiently enhanced satisfying the phase matching condition [27,28]. As a potential application, slow light-based all-optical switching/routing has been reported recently in an ultraslow light scheme [27]. This photon switching technique is used for the proof of the origin of the trapped light (discussed in Fig. 2(c)).

In the present Letter we experimentally demonstrate the ultralong trapped light using double spin coherence gratings. Unlike Ref. 9, whose light trapping is due to optical population gratings excited by a pair of counterpropagating coupling lights, the counterpropagating lights C and A in Fig. 1(c) have different frequencies resulting in no standing-wave gratings. Instead, the pair of resonant Raman optical fields, P and C, induces spin coherence moving gratings through the medium along the path of the light propagation [4,22-25]. As reported already a subsequent backward light A applied to the ultraslow light S induced by P and C generates phase conjugate PC, whose propagation direction is opposite to that of P. Then the lights A and PC can create another spin coherence moving gratings, but in an opposite direction. The forward ultraslow light S and the backward phase conjugate PC are coupled together and can be halted completely at a certain condition [24]:

$$\frac{\Omega_C}{g_C} = \frac{\Omega_A}{g_A}, \quad (1)$$

where $\Omega_i$ and $g_i$ are the Rabi frequency and coupling constant of light *i* with the atoms, respectively.

Figure 2 shows the result of the trapped light, where the trapping time is much longer than the expected value determined by the spin dephasing time $T^S_2{}^*$. The spin dephasing time is calculated from the inverse of the spin inhomogeneous width, $\Delta_S$=30 kHz: $T^S_2{}^* = 1/(\pi\Delta_S)$ [26], where the spin dephasing time is 10.6 μs. On the contrary the storage time in a spin homogeneously broadened medium [11-13] must be limited by the spin coherence relaxation time $T^S_2$. In Pr:YSO, $T^S_2$ ~ 500 μs [29]. In Fig. 2(a) an ultraslow light S is generated by P under C. When the backward coupling light A is turned on in Fig. 2(b), the ultraslow light S becomes trapped inside the medium until the light A is turned off [24]. As shown in Fig. 2(b), the trapped light (red dashed circle, S') comes out of (regenerates or accelerates) the medium when the light A is turned off. The intensity reduction of the regenerated (or accelerated) light S' compared to S is due mainly to the spin decoherence determined by the atom-field interactions. To prove the origin of the regenerated light S', a photon switching technique is applied [27]. In Fig. 2(c) the diffracted signal (red line) D represents the spin coherence depletion process, in which the spin coherence coexists with the regenerated light S' in Fig. 2(b), which originates at the slow light S in Fig. 2(a): For more information about the photon switching, please refer to Fig. 2 of Ref. 27. Thus, Fig. 2(c) demonstrates that the regenerated light S' comes directly from the ultraslow light S. The first peak which appeared immediately when the light A was turned on may indicate a leakage due to both the nonperfect geometry of the spin gratings and partial reflection on each spin grating node as mentioned already in the standing-wave grating method [9].

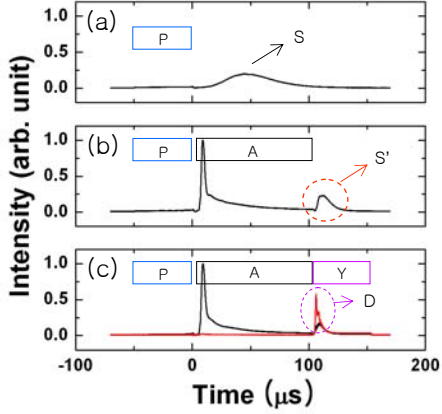

**Fig. 2.** (Color online) (a) Ultraslow light S under the action of the forward coupling light C only, (b) trapped light by adding a backward light A, and (c) photon switching of the trapped light as a proof of spin coherence excitation originated at the slow light S. Y is blue shifted by 1 MHz. The delay between P and A is 3 µs. The boxes P, A, and Y represent both the temporal position and pulse duration of each pulse. The control C covers all the light pulses (see Fig. 1(b)). The light power of R, P, C, A, and Y is 8, 0.9, 10, 20, and 20 mW, respectively. The diffraction signal is attenuated by a neutral density filter (ND 0.3).

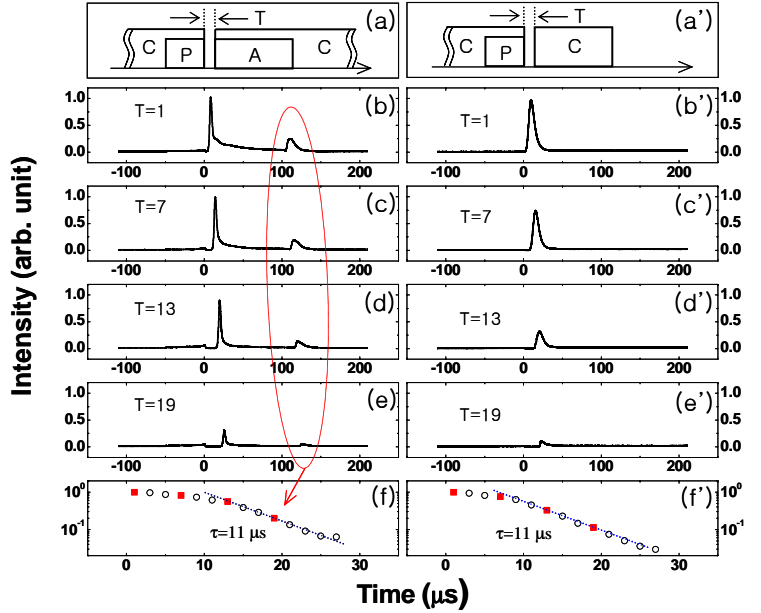

**Fig. 3.** (Color online) (a) and (a') Pulse sequence for trapped light and quantum memory, respectively, (b) − (e) and (b') − (e') Regenerated signal versus temporal delay T, (f) and (f') peak intensity of the regenerated light versus the delay T. For (b') − (e'), the peak power of C before (after) t = 0 is 10 (20) mW.

In Fig. 3, we analyze the trapping phenomenon observed in Fig. 2(b) using the quantum mapping mechanism [4,23]. According to the coherence conversion process [25,27], the magnitude of the regenerated (or retrieved) signal S' must be proportional to that of the spin coherence. The excited spin coherence by P and C (t < 0), however, begins to decay at t = 0 according to the overall spin dephasing time $T_2^{S*}$ at t=0. The left column of Fig. 3 shows stationary light for various dephasing conditions through control of delay time T (see Figs. 3(a) and 3(a')). The right column is given as a reference of quantum memory [11-13]. Figures 3(b) − 3(e) are counterparts of Figs. 3(b') – 3(e'). Figures 3(f) and 3(f') are the plots of the peak intensity of the regenerated light versus delay time T: Red squares are for Figs. 3(b) – 3(e) and for Figs. 3(b') – 3(e'), respectively. From Figs. 3(f) and 3(f'), the calculated decay time ($\tau$=11 µs) is exactly the same for each, which is very close to the value of the spin dephasing time $T_2^{S*}$ ($T_2^{S*} = 1/\pi\Delta^S = 10.6$ µs, where $\Delta^S$ is 30 kHz). Thus, we prove that the present method of stationary light (or trapping of light) is valid. This means that the medium's storage time cannot be longer than the overall spin dephasing time $T_2^{S*}$.

The inset of Fig. 4 shows intensities of the regenerated light S' in Fig. 2(b) for various pulse durations of the backward control field A: No Y is applied. In Fig. 4, each pulse intensity in the inset is plotted as a function of the pulse duration of A: The open squares are not shown in the inset. The best-fit curve for the data in Fig. 4 reveals a much longer decay time $\tau$ ($\tau$ = 588 µs) in comparison with that of Fig. 3(f), which is two orders of magnitude longer than the spin dephasing time $T_2^{S*}$ observed in Fig. 3:

$$I(t) = I_0 \exp(-t/\tau)^2. \qquad (2)$$

This value $\tau$ in Fig. 4 is comparable with the measured value of spin phase relaxation time $T_2^S$ (~500 µs) [29]. This means that the light trapping time of the present method in a spin inhomogeneously broadened medium is extended from $T_2^{S*}$ to $T_2^S$. If the intensity of A is weaker or stronger, then the decay time in Fig. 4 is much shortened because the balance condition is not satisfied (not shown). The outcome in Fig. 4 demonstrates that the present method of trapping light can be applied to ultralong trapped light even in an inhomogeneously broadened system. We understand that the ultralong trapped light is due to continuous interactions of photons with atoms as in the ultraslow light. However, the present method is completely different from the quantum mapping process [4,11-13], where the quantum mapping process cannot be used for nonlinear interactions during light storage. In addition to the enhanced nonlinear effect with the lengthened interaction time, use of wider bandwidth of spin transitions for shorter probe pulses must have an advantage to increase the delay bandwidth product [30], where the delay bandwidth product has been a critical limitation in slow-light-based quantum optical information processing [1,3-8].

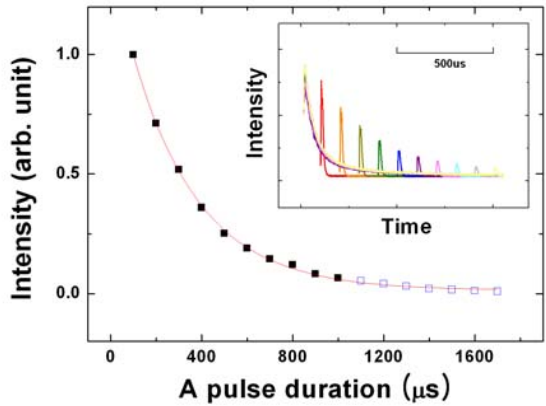

**Fig. 4.** (Color online) (a) Trapped light intensity for various pulse durations of the backward coupling A, and (b) intensity of the trapped light versus pulse duration of A. The solid line is a best-fit curve of Eq. (2).

In Conclusion, we have reported ultralong trapped light in an optically dense solid medium by using electromagnetically induced transparency, where the light trapping mechanism lies in two-photon induced spin coherence gratings completely different from standing-wave gratings or quantum mapping process. The trapping time of the halted light is two orders of magnitude longer than the expected value in the quantum mapping process. This observation opens a door to potential applications of light-matter interactions of ultralow power nonlinear optics as well as increased delay bandwidth product using a spin inhomogeneously broadened medium. Such nonlinear optics may also apply to quantum nondemolition measurement by greatly extending the interaction time, so that ultralow power light applications can be implemented.

Acknowledgment
BSH acknowledges that the present work was supported by the CRI program (Center for Photon Information Processing) of MEST via KOSEF, S. Korea. BSH also thanks S.A. Moiseev (University of Calgary, Canada) and M. S. Kim (Queen's University, UK) for helpful discussions.